\documentclass[twocolumn,prl,aps,floatfix,superscriptaddress]{revtex4-1}

\usepackage{color}
\usepackage{siunitx}
\usepackage{graphicx}
\usepackage{amsthm}
\usepackage{amsmath}
\usepackage{amssymb}
\usepackage{enumerate}

\DeclareMathOperator{\Tr}{Tr}

\newcommand{\ket}[1]{\vert #1 \rangle}

\newcommand{\bra}[1]{\langle #1 \vert}

\newcommand{\ren}{R\'{e}nyi~}
\usepackage[english]{babel}

\usepackage{tikz}
\newcommand*\circled[1]{\tikz[baseline=(char.base),text=white]{
            \node[shape=circle,fill=black,draw=black,inner sep=1pt] (char) {#1};}}

\begin{document}

\title{Entangling qubit registers via many-body states of ultracold atoms}

\author{R. G. Melko}
\affiliation{Department of Physics and Astronomy, University of Waterloo, Ontario, N2L 3G1, Canada}
\affiliation{Perimeter Institute for Theoretical Physics, Waterloo, Ontario N2L 2Y5, Canada}

\author{C. M. Herdman}
\affiliation{Department of Physics and Astronomy, University of Waterloo, Ontario, N2L 3G1, Canada}
\affiliation{Institute for Quantum Computing, University of Waterloo, Ontario, N2L 3G1, Canada}
\affiliation{Department of Chemistry,  University of Waterloo, Ontario, N2L 3G1, Canada}

\author{D. Iouchtchenko}
\affiliation{Department of Chemistry, University of Waterloo, Ontario, N2L 3G1, Canada}

\author{P.-N. Roy}
\affiliation{Department of Chemistry, University of Waterloo, Ontario, N2L 3G1, Canada}

\author{A. Del Maestro}
\email{Adrian.DelMaestro@uvm.edu}
\affiliation{Department of Physics, University of Vermont, Burlington, VT 05405, USA}

\begin{abstract}
Inspired by the experimental measurement of the \ren entanglement entropy in a
lattice of ultracold atoms by Islam \emph{et al.} [Nature {\bf 528}, 77
(2015)], we propose a method to entangle two spatially-separated qubits using
the quantum many-body state as a resource.  Through local operations accessible
in an experiment, entanglement is transferred to a qubit register
from atoms at the ends of a one-dimensional chain.  We
compute the {\it operational entanglement},
which bounds the entanglement physically transferable from the many-body
resource to the register, and discuss a protocol for its experimental
measurement.  Finally, we explore measures for the amount of entanglement
available in the register after transfer, suitable for use in quantum
information applications.
\end{abstract}

\maketitle


Islam \emph{et al.}~\cite{Islam:2049153} have performed a  measurement of the
\ren entanglement entropy in a one-dimensional optical lattice of ${}^{87}$Rb
atoms by exploiting a many-body analogue of the Hong-Ou-Mandel
\cite{Hong:1987gm} photon interference effect. After interfering two proximate
copies of an $L$-site lattice using the atomic control of a quantum gas
microscope \cite{Bakr:2010gd}, a measurement of the parity of the site resolved
particle occupation number provides access to the state overlap of the
two copies. If the initial copies are identical, this gives the purity of the 
state \cite{Daley:2012bd}.  Hence, if a globally pure state is partitioned into
spatial subregions, the many-body interference/parity measurement protocol
localized to a subregion yields the \ren entropy, a
measure of entanglement between subregions~\cite{Horodecki2009}.  This provides
an experimental probe of a remarkable feature of quantum mechanics with
no classical analogue: complete knowledge of the global state of a composite
system may not be enough to completely specify the state of a subsystem.

The advantage of measuring the \ren entropy as in
Ref.~[\onlinecite{Islam:2049153}] is that it encodes the entanglement
between subsystems in a scalar quantity that can be accessed through the
expectation values of local operators \cite{Daley:2012bd}.  This is in contrast
to other entanglement measures calculated directly from the full density
matrix, which is generally inaccessible in experiments without using full state
tomography \cite{James:2001bb}. In particular, there is currently no scalable
scheme for its reconstruction for $N$ interacting itinerant particles.
This fact makes the two-copy \ren entropy, $S_2(A) = -\log(\Tr\rho_A^2)$,
particularly well-suited for exploration in a quantum many-body system
bipartitioned into a spatial region $A$ and its complement $\bar{A}$.
 
$S_2$ has proved fruitful for the general characterization of many-body phases
and quantum phase transitions, \emph{e.g.}~through the exploration of its
scaling with subsystem size \cite{Amico:2008en}. Given that entanglement is a physical resource
that can be used for quantum information processing
\cite{Bennett1993,Vidal2003}, it is natural to ask whether this  
many-body entanglement can be harnessed for these tasks 
 \cite{Banchi2011,Yao2011a,Giampaolo2010,CamposVenuti:2007ku}. To
quantify this usable entanglement, one must take into account
physical restrictions that limit the amount of entanglement that may be
transferred to an external quantum register. For itinerant particles, a
super selection rule (SSR) due to particle number conservation provides one
key limitation \cite{Wiseman:2003jx}. Further restrictions are imposed if one
wants to entangle spatially separated qubits with only local operations on the
many-body system \cite{Horodecki:2000hr}. 

%
\begin{figure}[t]
\begin{center}
\includegraphics[width=1.0\columnwidth]{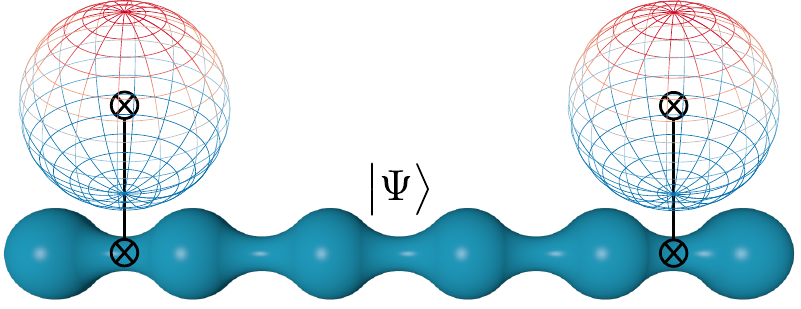}
\end{center}
\caption{(Color online) A schematic setup whereby entanglement can be
transferred from a quantum many-body state $|\Psi\rangle$ to a quantum register
composed of two spatially separated qubits (Bloch spheres).}
\label{fig:registers}
\end{figure}
%

In this paper we propose the general scheme shown in Fig.~\ref{fig:registers}
and present an experimental protocol, using the basic capabilities
of Islam \emph{et al.}~\cite{Islam:2049153}, to transfer some of the
entanglement in a many-body state of ultracold atoms to two spatially-separated
qubits composing a quantum register.  We emphasize the importance of the {\it
operational entanglement} as a bound on the transferable entanglement, and
discuss its measurement in the many-body state.  The demonstration of this
transfer would be proof of principle confirmation that a quantum register can be
entangled in current experimental apparatuses for ultracold atoms.

The ${}^{87}$Rb atoms of the Islam experiment
are confined to move in a deep one-dimensional optical lattice.
In their weakly interacting regime, the low energy dynamics of the atoms are
accurately governed by the lattice Bose-Hubbard Hamiltonian with $N$ particles
on $L$ sites:
\begin{equation}
H = -J \sum_{i=1}^{L-1} \left(b_{i}^\dag b_{i+1}^{\phantom \dag} + \text{h.c.}\right)
     + \frac{U}{2} \sum_{i=1}^L n_i\left(n_i -1\right),
\label{eq:BHHam}
\end{equation}
where $b_i^\dag (b_i^{\phantom \dag})$ creates (annihilates) a boson, and $n_i
= b_{i}^\dag b_{i}^{\phantom \dag}$ counts the number of atoms on site $i$.
$J$ sets the rate of tunneling between sites while $U$ parametrizes the strength of
the on-site repulsion between atoms. In an experiment, the interaction strength
between ${}^{87}$Rb atoms is fixed by their $s$-wave scattering length, while $J$
can be tuned by manipulating the height of the optical lattice.  In the
thermodynamic limit at unit filling ($N=L$), Eq.~\eqref{eq:BHHam} exhibits two distinct
phases: a Mott insulator for $U/J \gg 1$ and a superfluid for $U/J
\ll 1$, both of which are observed experimentally.  A quantum phase transition
separates these two phases at $(U/J)_c \approx 3.3$ \cite{Cazalilla:2011dma,
Carrasquilla:2013dg, Boeris:2015un, Astrakharchik:2015ww}.

The spatially delocalized nature of particles in the superfluid phase suggests
that it should be significantly more entangled under a spatial bipartition 
than a Mott insulator with localized particles. This is manifest as an 
increase in $S_2$ accompanying the onset of delocalization at $U/J \sim
\mathrm{O}(1)$ observed in the experiment for $N=4$ atoms \cite{Islam:2049153}.
The same experimental capabilities that allow for the measurement of the
entanglement in an optical lattice can also be used to transfer entanglement to
spatially separated qubits, that can be employed as a quantum register for
 information processing tasks via logic gates.  This entanglement
transfer procedure is limited by the SSR that forbids the creation of a
coherent superposition of states with different local particle
number \cite{Wiseman:2003jx}. Thus entanglement that arises \emph{solely} due to
particle fluctuations between subregions is not physically transferable to a
register without a global phase reference \cite{Aharonov:1967be}.

To address this issue, Wiseman and Vaccaro \cite{Wiseman:2003jx} introduced the
concept of \emph{operational entanglement} -- the amount of entanglement that
can be extracted from a resource (many-body state) and transferred to a quantum
register in the presence of a SSR.  Conceptually it is the weighted sum of the
spatial entanglement when projecting onto states of fixed local particle
number. For the two copy \ren entropy it is defined as:
\begin{equation}
    S_2^{\rm op}(A) = \sum_n P_n S_2\left(A_n\right),
\label{eq:S2op}
\end{equation}
where $S_2(A_n)$ is the \ren entropy evaluated for the reduced density
matrix $\rho_{A_n} = \hat{P}_n \rho_A \hat{P}_n/P_n$ projected by $\hat{P}_n$
onto states of fixed local particle number $n$ in subsystem $A$.  The summation
is over all possible local particle number states in the subregion with
$n=0,\ldots,N$, each having probability $P_n = \bra{\Psi} \hat{P}_n
\ket{\Psi}$.  This
projection is a local operation that can only decrease entanglement
\cite{Horodecki:2000hr} so $S_2^{\rm op} \le S_2$. 

Thus it is $S_2^{\rm op}$, not $S_2$ which bounds the amount of entanglement
that can be generated in the register using local operations and classical
communication (LOCC).  A measurement of $S_2^{\rm op}$ is
possible with a simple modification of the experimental interference/parity measurement
procedure \cite{Daley:2012bd,Islam:2049153}. This requires that a projection
onto states of definite subsystem particle number $n$ be made after
interference, despite the fact that this may mix particles between copies.
Fortunately, one can show that after an ensemble average is taken, only cases
with equal particle number in region $A$ of both replicas will contribute to
the expectation value of the parity operator.  Therefore, by simply collecting
$n$-resolved statistics of the \ren entropy, Eq.~\eqref{eq:S2op} can be used to
experimentally measure the operational entanglement. 

%
\begin{figure}[t]
\begin{center}
\includegraphics[width=1.0\columnwidth]{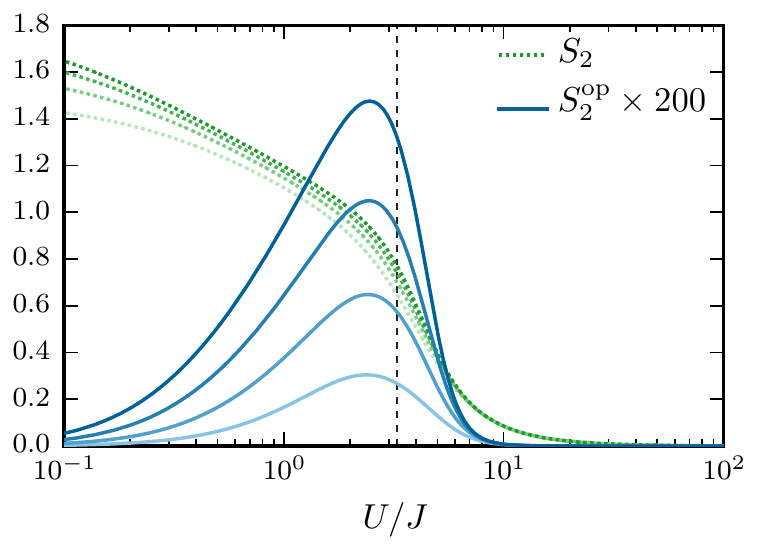}
\end{center}
\caption{(Color online) The spatial second \ren entropy $S_2$ and operational
entanglement $S_2^{\rm op}$ for symmetric bipartitions $\ell=L/2$ of the
Bose-Hubbard model. Curves increase in saturation for $L = N = 6, 8, 10, 12$.
The dashed vertical line indicates the location of the thermodynamic phase
transition. }
\label{fig:entropies_sym}
\end{figure}
%

In order to explore which parameter regime maximizes operational entanglement,
we calculate $S_2^{\rm op}$ in the Bose-Hubbard model.  Experiments on
${}^{87}$Rb in the near future should be possible with $4<N
\lesssim 10$, and we study the ground states of systems with sizes of this
order via exact numerical diagonalization of Eq.~\eqref{eq:BHHam}.  In
Fig.~\ref{fig:entropies_sym}, we compare the two-copy \ren entropy for a
symmetric spatial bipartition to the operational entanglement for a range of
$U/J$ and $N$ relevant for experiment.  Unlike the entropy under a spatial
bipartition, which is maximum deep in the superfluid phase (or the particle
entanglement, which is maximum deep in the Mott phase \cite{Herdman:2014ey}),
$S_2^{\rm op}$ displays a peak at an intermediate value of the interaction.
While for these system sizes, the peak is not positioned directly at the
thermodynamic-limit critical point $(U/J)_c \approx 3.3$, it appears to
approach this value as the system size is increased.  This suggests that
the appropriate experimental parameters for maximizing the transfer of
many-body entanglement to a system of quantum registers will be those that tune
the system to near the superfluid-Mott transition.

As seen in Fig.~\ref{fig:entropies_sym}, $S_2^{\rm op}$ is necessarily smaller
than $S_2$, as it does not include entanglement generated by particle
fluctuations between subsystems that is not physically accessible due to the
SSR. Additionally, $S_2^{\rm op}$ is reduced
as interactions in Eq.~\eqref{eq:BHHam} are strictly onsite and occur at
fixed subsystem occupation through 2nd order processes.  Thus, the behavior of
the physically accessible entanglement differs from $S_2$ 
both qualitatively and quantitatively.

\begin{figure}[t]
\begin{center}
\includegraphics[width=1.0\columnwidth]{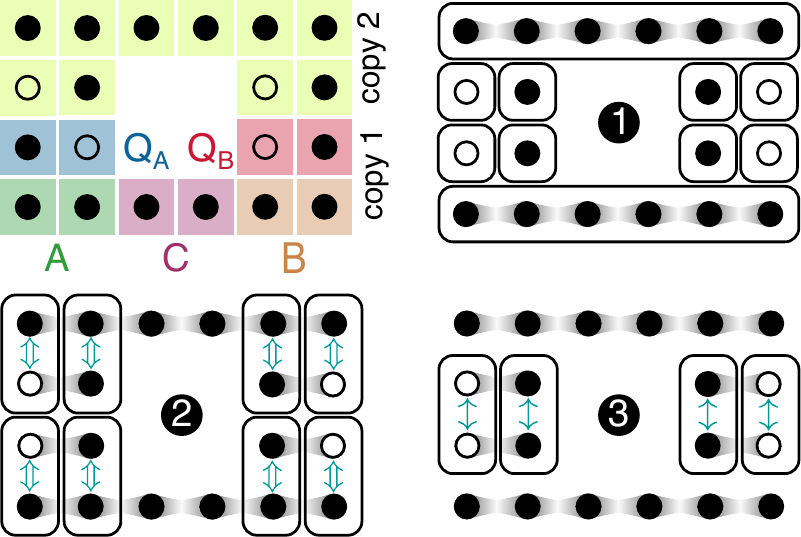}
\end{center}
\caption{(Color online) Upper left: an array of 20 optical lattice sites 
forming the two copies necessary to measure the second \ren
entropy.  Remaining panels: the protocol (described
in the text) to transfer entanglement from a many-body state in $A\cup B
\cup C$ to spatially separated qubits $Q_A$ and $Q_B$. Solid lines
correspond to a large tunnel barrier, double arrows represent the application
of a SWAP operation and single arrows indicate performing many-body
interference.}
\label{fig:exp_procedure}
\end{figure}

Given that the operational entanglement indicates that some of the entanglement
between spatial subregions of the many-body ground state may be transferred to 
an external quantum register using LOCC, we now describe an experimental procedure 
to do so. This allows the many-body state to act as an entanglement resource for 
quantum information protocols.  We concentrate on the minimal $L=N=6$ Bose-Hubbard
system where entanglement my be transferred to two spatially separated qubits.
Each qubit is comprised of one atom occupying one of two neighboring lattice
sites adjacent to the Bose-Hubbard chain; the two locations of the atom
provide the logical states.  Thus, the physical system we describe
consists of 10 total lattice sites, which must be doubled as shown in
Fig.~\ref{fig:exp_procedure} if a two-copy \ren measurement is to be made on
the final entanglement between the qubits.

%
%

The starting point is the isolation of a $6\times 4$ array of atoms that can be
prepared deep in the Mott phase.  This array
includes the many-body entanglement resource, which will be partitioned into
three spatial subregions with two sites each (labeled $A,B,C$), two qubit
registers, $Q_A$ and $Q_B$, and a copy that will be employed to read out the
amount of entanglement generated between $Q_A$ and $Q_B$. To manipulate and
 measure entanglement in the system, we define a pairwise hopping unitary operator:
 \begin{equation}
     U_{ij} \left( \phi \right) \equiv \exp\left[ i \phi \left( b_i^\dagger b_j
     + {\rm h.c.} \right) \right].
 \end{equation}
 This is a trivial generalization of the beam-splitter operation reported in
 Ref.~[\onlinecite{Islam:2049153}] (where $\phi = \pi/4$) and $\phi = \pi/2$
 corresponds to a SWAP gate between $i$ and $j$ within the $n_{i,j}= 0,1$
 subspace. Additionally this physical operation can be used to perform single
 qubit rotations when applied within a single qubit. As $U_{ij}$ will not
 generally preserve particle number within the resource and qubits (and thus
 not remain in the logical subspace of the qubits), subsystem resolved
 particle occupation number measurements must be used to post-select states
 that have exactly one particle in each of $A$ and $B$. 
 
Transfer of many-body entanglement to the register and its subsequent
measurement can be accomplished via the three step procedure depicted in
Fig.~\ref{fig:exp_procedure}.  \circled{1} The optical lattice within the array
is manipulated such that large barriers (as indicated by solid lines) isolate 
the many-body resource. Each qubit must be constructed with
exactly one particle between its two sites, with the barrier between them
remaining high throughout the experiment.  The many-body resource can be
prepared identically to Ref.~[\onlinecite{Islam:2049153}] with the lattice
strength tuned near the critical value $(U/J)_c$ to maximize the operational
entanglement as discussed above.  \circled{2} A SWAP operation (double arrow)
is performed between $A\Leftrightarrow Q_A$ by applying the unitary hopping
operator $U_{1,1'}(\pi/2)U_{2,2'}(\pi/2)$, where sites $1,2$ are in region $A$,
while $1',2'$ label adjacent sites in $Q_A$. This is repeated for
$B\Leftrightarrow Q_B$ and the identical procedure is performed in the copy.
Thus entanglement is transferred from the many-body resource to the spatially
separated qubits.  \circled{3} To read out this entanglement, a beam-splitter
operation (single arrow) is performed between the two copies of $Q_A$ and
$Q_B$, followed by a subsystem resolved particle number measurement where
instances with one atom in each qubit  are post-selected. 

The above procedure will transfer many-body entanglement to a quantum register.
As only $A$ and $B$ are swapped with the register, its density matrix
$\rho_{Q_A,Q_B}$ will generically be in a mixed state, even if the initial
many-body state ($\rho_{ABC}$) was pure. Consequently, the mutual information
$I_2({AB}) = S_2(A) + S_2(B) - S_2(AB)$
will have contributions from both
classical correlations and quantum entanglement.   $I_2(AB)$
is measurable in current experiments combined with post-selection to conserve
particle number in $Q_{A/B}$.

To quantify only the desired generation of quantum entanglement between the
qubits, we compute various measures of mixed state entanglement for the reduced
density matrix $\rho_{AB}$ of the many-body ground state. Unlike for pure
states, where the von Neumann entropy is the unique and
appropriate entanglement measure, for mixed states, there are a
variety of entanglement measures with different physical meanings. For example,
the entanglement of formation $E_F$, roughly defined as the amount of
entanglement required to form the mixed state, can be directly computed for
any two qubit density matrix \cite{Wootters1998}.  The logarithmic negativity
$E_\mathcal{N}$ depends on the sum of the negative eigenvalues of the
density matrix after a partial transpose, and thus is readily computable for any
density matrix \cite{Vidal2002}. It  provides an upper bound to the amount of
entanglement that can be extracted from the mixed state using LOCC.
 
%
\begin{figure}[t] \begin{center}
\includegraphics[width=1.0\columnwidth]{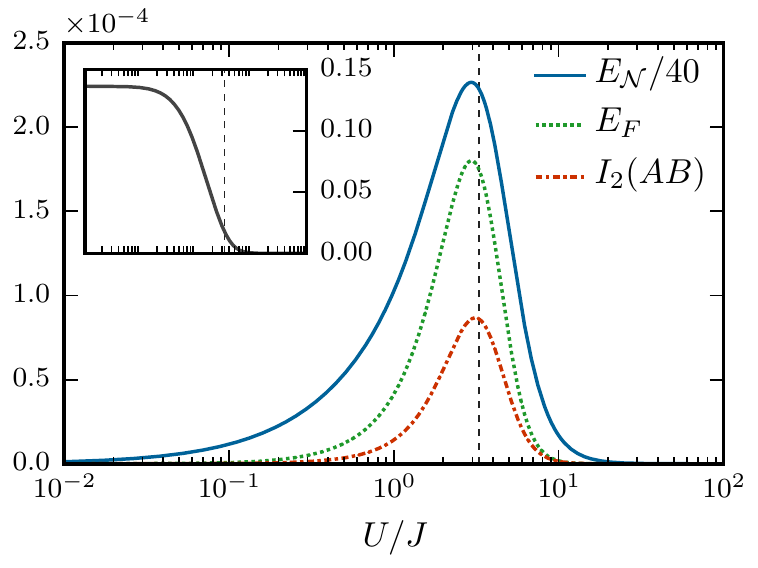} \end{center}
\caption{(Color online) The logarithmic negativity $E_\mathcal{N}$,
    entanglement of formation $E_F$, and mutual information $I_2({AB})$ of the
    spatially separated qubits obtained from the $L = N = 6$ Bose-Hubbard
    ground state. The inset shows the probability of projecting onto a state
    with a single particle in each of $A$ and $B$.  The dashed vertical
    line indicates the location of the thermodynamic phase transition. }
\label{fig:entropies_qubit} \end{figure}
%
In Fig.~\ref{fig:entropies_qubit} we have plotted $I_2({AB})$, $E_F$, and
$E_\mathcal{N}$ of $\rho_{AB}$ for the ground state of Eq.~\eqref{eq:BHHam} in
the 6-site geometry of Fig.~\ref{fig:exp_procedure}, projected onto states with
a single particle occupying $A$ and $B$.  We find that all these measures peak
near the quantum phase transition
\cite{Osterloh2002,Osborne2002,Frerot:2015tl}. The peaks coincide with the
parameter region of maximal operational entanglement desired for optimal
transfer between resource and register.  $E_{\mathcal{N}}>0$ is a necessary and
sufficient condition for a two qubit state to be inseparable
\cite{Horodecki1996a} such that it can be distilled to form a maximally
entangled state \cite{Horodecki1997}. This implies that near the critical point
the many-body resource has entanglement that can be extracted and distilled.
Although there is no general relationship between $I_2$ and the entanglement
measures $E_F$ and $E_{\mathcal{N}}$, in this case we can compute the
relationship exactly for the Bose-Hubbard model.  Thus, measurement of $I_2$ in
an experimental regime where the Bose-Hubbard parameters are known will
provide an estimate of the entanglement that can be generated between the
qubits through the relationship calculated in  Fig.~\ref{fig:entropies_qubit}.

In conclusion, we have introduced an experimental procedure for the transfer of 
entanglement from a many-body resource to spatially separated qubits forming a
register suitable for quantum information processing.
Conservation of particle number limits the amount of entanglement transferable from the resource,
as quantified by the operational entanglement.
The precise control of the current generation of quantum emulator experiments 
enables the faithful creation of lattice Bose-Hubbard models using ultracold atoms.
This allows us to quantify the operational entanglement using exact calculations,
and we find that the transferable entanglement is maximized near the quantum
phase transition between the Mott insulator and superfluid phases.
This is in contrast to the naive expectation that transfer should occur in the
superfluid phase, where experiments have confirmed that the two-copy \ren
entanglement is largest \cite{Islam:2049153}.

We have introduced a measurement protocol to experimentally probe the
entanglement transferred by this procedure that employs a variation of a
many-body interference technique \cite{Islam:2049153,Daley:2012bd}.  It is explicitly
described for the transfer of entanglement from a 6-site resource to a register
composed of two 2-site qubits --  20 lattice sites in total. It can be easily scaled
to arbitrary size as experimental technology progresses.  Our Bose-Hubbard
calculations quantify the relationship between a mutual information accessible
by this protocol and well-known measures for entanglement in mixed states.

The ability to engineer a wealth of variations of the Bose-Hubbard model will
open up exciting prospects for extensions and optimizations of our results,
through inhomogeneous parameters, topologies, and dimensionality.  The
experimental implementation of our protocol will demonstrate the potential of
using many-body states of ultracold atoms as an entanglement resource for
quantum information processing.

This work would not have been possible without discussions with R. Islam and A.
Kaufmann. We thank J. Carrasquilla for his insights into the 1D Bose-Hubbard
model and A. Brodutch for discussions about entanglement in mixed states. This research was supported by NSERC of Canada, the
Canada Research Chair Program, the Perimeter Institute for Theoretical Physics
(PI) and the National Science Foundation under Grant
No.~NSF~PHY11-25915.  Research at PI is supported by the Government of
Canada through Industry Canada and by the Province of Ontario through the
Ministry of Economic Development \& Innovation.

\bibliography{bh_refs}

\begin{thebibliography}{28}%
\makeatletter
\providecommand \@ifxundefined [1]{%
 \@ifx{#1\undefined}
}%
\providecommand \@ifnum [1]{%
 \ifnum #1\expandafter \@firstoftwo
 \else \expandafter \@secondoftwo
 \fi
}%
\providecommand \@ifx [1]{%
 \ifx #1\expandafter \@firstoftwo
 \else \expandafter \@secondoftwo
 \fi
}%
\providecommand \natexlab [1]{#1}%
\providecommand \enquote  [1]{``#1''}%
\providecommand \bibnamefont  [1]{#1}%
\providecommand \bibfnamefont [1]{#1}%
\providecommand \citenamefont [1]{#1}%
\providecommand \href@noop [0]{\@secondoftwo}%
\providecommand \href [0]{\begingroup \@sanitize@url \@href}%
\providecommand \@href[1]{\@@startlink{#1}\@@href}%
\providecommand \@@href[1]{\endgroup#1\@@endlink}%
\providecommand \@sanitize@url [0]{\catcode `\\12\catcode `\$12\catcode
  `\&12\catcode `\#12\catcode `\^12\catcode `\_12\catcode `\%12\relax}%
\providecommand \@@startlink[1]{}%
\providecommand \@@endlink[0]{}%
\providecommand \url  [0]{\begingroup\@sanitize@url \@url }%
\providecommand \@url [1]{\endgroup\@href {#1}{\urlprefix }}%
\providecommand \urlprefix  [0]{URL }%
\providecommand \Eprint [0]{\href }%
\providecommand \doibase [0]{http://dx.doi.org/}%
\providecommand \selectlanguage [0]{\@gobble}%
\providecommand \bibinfo  [0]{\@secondoftwo}%
\providecommand \bibfield  [0]{\@secondoftwo}%
\providecommand \translation [1]{[#1]}%
\providecommand \BibitemOpen [0]{}%
\providecommand \bibitemStop [0]{}%
\providecommand \bibitemNoStop [0]{.\EOS\space}%
\providecommand \EOS [0]{\spacefactor3000\relax}%
\providecommand \BibitemShut  [1]{\csname bibitem#1\endcsname}%
\let\auto@bib@innerbib\@empty
\bibitem [{\citenamefont {Islam}\ \emph {et~al.}(2015)\citenamefont {Islam},
  \citenamefont {Ma}, \citenamefont {Preiss}, \citenamefont {Tai},
  \citenamefont {Lukin}, \citenamefont {Rispoli},\ and\ \citenamefont
  {Greiner}}]{Islam:2049153}%
  \BibitemOpen
  \bibfield  {author} {\bibinfo {author} {\bibfnamefont {R.}~\bibnamefont
  {Islam}}, \bibinfo {author} {\bibfnamefont {R.}~\bibnamefont {Ma}}, \bibinfo
  {author} {\bibfnamefont {P.~M.}\ \bibnamefont {Preiss}}, \bibinfo {author}
  {\bibfnamefont {M.~E.}\ \bibnamefont {Tai}}, \bibinfo {author} {\bibfnamefont
  {A.}~\bibnamefont {Lukin}}, \bibinfo {author} {\bibfnamefont
  {M.}~\bibnamefont {Rispoli}}, \ and\ \bibinfo {author} {\bibfnamefont
  {M.}~\bibnamefont {Greiner}},\ }\href {\doibase 10.1038/nature15750}
  {\bibfield  {journal} {\bibinfo  {journal} {Nature}\ }\textbf {\bibinfo
  {volume} {528}},\ \bibinfo {pages} {77} (\bibinfo {year} {2015})}\BibitemShut
  {NoStop}%
\bibitem [{\citenamefont {Hong}\ \emph {et~al.}(1987)\citenamefont {Hong},
  \citenamefont {Ou},\ and\ \citenamefont {Mandel}}]{Hong:1987gm}%
  \BibitemOpen
  \bibfield  {author} {\bibinfo {author} {\bibfnamefont {C.~K.}\ \bibnamefont
  {Hong}}, \bibinfo {author} {\bibfnamefont {Z.~Y.}\ \bibnamefont {Ou}}, \ and\
  \bibinfo {author} {\bibfnamefont {L.}~\bibnamefont {Mandel}},\ }\href
  {\doibase 10.1103/PhysRevLett.59.2044} {\bibfield  {journal} {\bibinfo
  {journal} {Phys. Rev. Lett.}\ }\textbf {\bibinfo {volume} {59}},\ \bibinfo
  {pages} {2044} (\bibinfo {year} {1987})}\BibitemShut {NoStop}%
\bibitem [{\citenamefont {Bakr}\ \emph {et~al.}(2010)\citenamefont {Bakr},
  \citenamefont {Peng}, \citenamefont {Tai}, \citenamefont {Ma}, \citenamefont
  {Simon}, \citenamefont {Gillen}, \citenamefont {F{\"o}lling}, \citenamefont
  {Pollet},\ and\ \citenamefont {Greiner}}]{Bakr:2010gd}%
  \BibitemOpen
  \bibfield  {author} {\bibinfo {author} {\bibfnamefont {W.~S.}\ \bibnamefont
  {Bakr}}, \bibinfo {author} {\bibfnamefont {A.}~\bibnamefont {Peng}}, \bibinfo
  {author} {\bibfnamefont {M.~E.}\ \bibnamefont {Tai}}, \bibinfo {author}
  {\bibfnamefont {R.}~\bibnamefont {Ma}}, \bibinfo {author} {\bibfnamefont
  {J.}~\bibnamefont {Simon}}, \bibinfo {author} {\bibfnamefont {J.~I.}\
  \bibnamefont {Gillen}}, \bibinfo {author} {\bibfnamefont {S.}~\bibnamefont
  {F{\"o}lling}}, \bibinfo {author} {\bibfnamefont {L.}~\bibnamefont {Pollet}},
  \ and\ \bibinfo {author} {\bibfnamefont {M.}~\bibnamefont {Greiner}},\ }\href
  {\doibase 10.1126/science.1192368} {\bibfield  {journal} {\bibinfo  {journal}
  {Science}\ }\textbf {\bibinfo {volume} {329}},\ \bibinfo {pages} {547}
  (\bibinfo {year} {2010})}\BibitemShut {NoStop}%
\bibitem [{\citenamefont {Daley}\ \emph {et~al.}(2012)\citenamefont {Daley},
  \citenamefont {Pichler}, \citenamefont {Schachenmayer},\ and\ \citenamefont
  {Zoller}}]{Daley:2012bd}%
  \BibitemOpen
  \bibfield  {author} {\bibinfo {author} {\bibfnamefont {A.~J.}\ \bibnamefont
  {Daley}}, \bibinfo {author} {\bibfnamefont {H.}~\bibnamefont {Pichler}},
  \bibinfo {author} {\bibfnamefont {J.}~\bibnamefont {Schachenmayer}}, \ and\
  \bibinfo {author} {\bibfnamefont {P.}~\bibnamefont {Zoller}},\ }\href
  {\doibase 10.1103/PhysRevLett.109.020505} {\bibfield  {journal} {\bibinfo
  {journal} {Phys. Rev. Lett.}\ }\textbf {\bibinfo {volume} {109}},\ \bibinfo
  {pages} {020505} (\bibinfo {year} {2012})}\BibitemShut {NoStop}%
\bibitem [{\citenamefont {Horodecki}\ \emph {et~al.}(2009)\citenamefont
  {Horodecki}, \citenamefont {Horodecki},\ and\ \citenamefont
  {Horodecki}}]{Horodecki2009}%
  \BibitemOpen
  \bibfield  {author} {\bibinfo {author} {\bibfnamefont {R.}~\bibnamefont
  {Horodecki}}, \bibinfo {author} {\bibfnamefont {M.}~\bibnamefont
  {Horodecki}}, \ and\ \bibinfo {author} {\bibfnamefont {K.}~\bibnamefont
  {Horodecki}},\ }\href {\doibase 10.1103/RevModPhys.81.865} {\bibfield
  {journal} {\bibinfo  {journal} {Rev. Mod. Phys.}\ }\textbf {\bibinfo {volume}
  {81}},\ \bibinfo {pages} {865} (\bibinfo {year} {2009})}\BibitemShut
  {NoStop}%
\bibitem [{\citenamefont {James}\ \emph {et~al.}(2001)\citenamefont {James},
  \citenamefont {Kwiat}, \citenamefont {Munro},\ and\ \citenamefont
  {White}}]{James:2001bb}%
  \BibitemOpen
  \bibfield  {author} {\bibinfo {author} {\bibfnamefont {D.~F.~V.}\
  \bibnamefont {James}}, \bibinfo {author} {\bibfnamefont {P.~G.}\ \bibnamefont
  {Kwiat}}, \bibinfo {author} {\bibfnamefont {W.~J.}\ \bibnamefont {Munro}}, \
  and\ \bibinfo {author} {\bibfnamefont {A.~G.}\ \bibnamefont {White}},\ }\href
  {\doibase 10.1103/PhysRevA.64.052312} {\bibfield  {journal} {\bibinfo
  {journal} {Phys. Rev. A}\ }\textbf {\bibinfo {volume} {64}},\ \bibinfo
  {pages} {052312} (\bibinfo {year} {2001})}\BibitemShut {NoStop}%
\bibitem [{\citenamefont {Amico}\ \emph {et~al.}(2008)\citenamefont {Amico},
  \citenamefont {Osterloh},\ and\ \citenamefont {Vedral}}]{Amico:2008en}%
  \BibitemOpen
  \bibfield  {author} {\bibinfo {author} {\bibfnamefont {L.}~\bibnamefont
  {Amico}}, \bibinfo {author} {\bibfnamefont {A.}~\bibnamefont {Osterloh}}, \
  and\ \bibinfo {author} {\bibfnamefont {V.}~\bibnamefont {Vedral}},\ }\href
  {\doibase 10.1103/RevModPhys.80.517} {\bibfield  {journal} {\bibinfo
  {journal} {Rev. Mod. Phys.}\ }\textbf {\bibinfo {volume} {80}},\ \bibinfo
  {pages} {517} (\bibinfo {year} {2008})}\BibitemShut {NoStop}%
\bibitem [{\citenamefont {Bennett}\ \emph {et~al.}(1993)\citenamefont
  {Bennett}, \citenamefont {Brassard}, \citenamefont {Cr{\'{e}}peau},
  \citenamefont {Jozsa}, \citenamefont {Peres},\ and\ \citenamefont
  {Wootters}}]{Bennett1993}%
  \BibitemOpen
  \bibfield  {author} {\bibinfo {author} {\bibfnamefont {C.~H.}\ \bibnamefont
  {Bennett}}, \bibinfo {author} {\bibfnamefont {G.}~\bibnamefont {Brassard}},
  \bibinfo {author} {\bibfnamefont {C.}~\bibnamefont {Cr{\'{e}}peau}}, \bibinfo
  {author} {\bibfnamefont {R.}~\bibnamefont {Jozsa}}, \bibinfo {author}
  {\bibfnamefont {A.}~\bibnamefont {Peres}}, \ and\ \bibinfo {author}
  {\bibfnamefont {W.~K.}\ \bibnamefont {Wootters}},\ }\href {\doibase
  10.1103/PhysRevLett.70.1895} {\bibfield  {journal} {\bibinfo  {journal}
  {Phys. Rev. Lett.}\ }\textbf {\bibinfo {volume} {70}},\ \bibinfo {pages}
  {1895} (\bibinfo {year} {1993})}\BibitemShut {NoStop}%
\bibitem [{\citenamefont {Vidal}(2003)}]{Vidal2003}%
  \BibitemOpen
  \bibfield  {author} {\bibinfo {author} {\bibfnamefont {G.}~\bibnamefont
  {Vidal}},\ }\href {\doibase 10.1103/PhysRevLett.91.147902} {\bibfield
  {journal} {\bibinfo  {journal} {Phys. Rev. Lett.}\ }\textbf {\bibinfo
  {volume} {91}},\ \bibinfo {pages} {147902} (\bibinfo {year} {2003})},\
  \Eprint {http://arxiv.org/abs/0301063} {0301063} \BibitemShut {NoStop}%
\bibitem [{\citenamefont {Banchi}\ \emph {et~al.}(2011)\citenamefont {Banchi},
  \citenamefont {Bayat}, \citenamefont {Verrucchi},\ and\ \citenamefont
  {Bose}}]{Banchi2011}%
  \BibitemOpen
  \bibfield  {author} {\bibinfo {author} {\bibfnamefont {L.}~\bibnamefont
  {Banchi}}, \bibinfo {author} {\bibfnamefont {A.}~\bibnamefont {Bayat}},
  \bibinfo {author} {\bibfnamefont {P.}~\bibnamefont {Verrucchi}}, \ and\
  \bibinfo {author} {\bibfnamefont {S.}~\bibnamefont {Bose}},\ }\href {\doibase
  10.1103/PhysRevLett.106.140501} {\bibfield  {journal} {\bibinfo  {journal}
  {Phys. Rev. Lett.}\ }\textbf {\bibinfo {volume} {106}},\ \bibinfo {pages}
  {140501} (\bibinfo {year} {2011})}\BibitemShut {NoStop}%
\bibitem [{\citenamefont {Yao}\ \emph {et~al.}(2011)\citenamefont {Yao},
  \citenamefont {Jiang}, \citenamefont {Gorshkov}, \citenamefont {Gong},
  \citenamefont {Zhai}, \citenamefont {Duan},\ and\ \citenamefont
  {Lukin}}]{Yao2011a}%
  \BibitemOpen
  \bibfield  {author} {\bibinfo {author} {\bibfnamefont {N.~Y.}\ \bibnamefont
  {Yao}}, \bibinfo {author} {\bibfnamefont {L.}~\bibnamefont {Jiang}}, \bibinfo
  {author} {\bibfnamefont {a.~V.}\ \bibnamefont {Gorshkov}}, \bibinfo {author}
  {\bibfnamefont {Z.-X.}\ \bibnamefont {Gong}}, \bibinfo {author}
  {\bibfnamefont {A.}~\bibnamefont {Zhai}}, \bibinfo {author} {\bibfnamefont
  {L.-M.}\ \bibnamefont {Duan}}, \ and\ \bibinfo {author} {\bibfnamefont
  {M.~D.}\ \bibnamefont {Lukin}},\ }\href {\doibase
  10.1103/PhysRevLett.106.040505} {\bibfield  {journal} {\bibinfo  {journal}
  {Phys. Rev. Lett.}\ }\textbf {\bibinfo {volume} {106}},\ \bibinfo {pages}
  {040505} (\bibinfo {year} {2011})}\BibitemShut {NoStop}%
\bibitem [{\citenamefont {Giampaolo}\ and\ \citenamefont
  {Illuminati}(2010)}]{Giampaolo2010}%
  \BibitemOpen
  \bibfield  {author} {\bibinfo {author} {\bibfnamefont {S.~M.}\ \bibnamefont
  {Giampaolo}}\ and\ \bibinfo {author} {\bibfnamefont {F.}~\bibnamefont
  {Illuminati}},\ }\href {\doibase 10.1088/1367-2630/12/2/025019} {\bibfield
  {journal} {\bibinfo  {journal} {New J. Phys.}\ }\textbf {\bibinfo {volume}
  {12}},\ \bibinfo {pages} {025019} (\bibinfo {year} {2010})}\BibitemShut
  {NoStop}%
\bibitem [{\citenamefont {Campos~Venuti}\ \emph {et~al.}(2007)\citenamefont
  {Campos~Venuti}, \citenamefont {Giampaolo}, \citenamefont {Illuminati},\ and\
  \citenamefont {Zanardi}}]{CamposVenuti:2007ku}%
  \BibitemOpen
  \bibfield  {author} {\bibinfo {author} {\bibfnamefont {L.}~\bibnamefont
  {Campos~Venuti}}, \bibinfo {author} {\bibfnamefont {S.~M.}\ \bibnamefont
  {Giampaolo}}, \bibinfo {author} {\bibfnamefont {F.}~\bibnamefont
  {Illuminati}}, \ and\ \bibinfo {author} {\bibfnamefont {P.}~\bibnamefont
  {Zanardi}},\ }\href {\doibase 10.1103/PhysRevA.76.052328} {\bibfield
  {journal} {\bibinfo  {journal} {Phys. Rev. A}\ }\textbf {\bibinfo {volume}
  {76}},\ \bibinfo {pages} {052328} (\bibinfo {year} {2007})}\BibitemShut
  {NoStop}%
\bibitem [{\citenamefont {Wiseman}\ and\ \citenamefont
  {Vaccaro}(2003)}]{Wiseman:2003jx}%
  \BibitemOpen
  \bibfield  {author} {\bibinfo {author} {\bibfnamefont {H.~M.}\ \bibnamefont
  {Wiseman}}\ and\ \bibinfo {author} {\bibfnamefont {J.~A.}\ \bibnamefont
  {Vaccaro}},\ }\href {\doibase 10.1103/PhysRevLett.91.097902} {\bibfield
  {journal} {\bibinfo  {journal} {Phys. Rev. Lett.}\ }\textbf {\bibinfo
  {volume} {91}},\ \bibinfo {pages} {097902} (\bibinfo {year}
  {2003})}\BibitemShut {NoStop}%
\bibitem [{\citenamefont {Horodecki}\ \emph {et~al.}(2000)\citenamefont
  {Horodecki}, \citenamefont {Horodecki},\ and\ \citenamefont
  {Horodecki}}]{Horodecki:2000hr}%
  \BibitemOpen
  \bibfield  {author} {\bibinfo {author} {\bibfnamefont {M.}~\bibnamefont
  {Horodecki}}, \bibinfo {author} {\bibfnamefont {P.}~\bibnamefont
  {Horodecki}}, \ and\ \bibinfo {author} {\bibfnamefont {R.}~\bibnamefont
  {Horodecki}},\ }\href {\doibase 10.1103/PhysRevLett.84.2014} {\bibfield
  {journal} {\bibinfo  {journal} {Phys. Rev. Lett.}\ }\textbf {\bibinfo
  {volume} {84}},\ \bibinfo {pages} {2014} (\bibinfo {year}
  {2000})}\BibitemShut {NoStop}%
\bibitem [{\citenamefont {Cazalilla}\ \emph {et~al.}(2011)\citenamefont
  {Cazalilla}, \citenamefont {Citro}, \citenamefont {Giamarchi}, \citenamefont
  {Orignac},\ and\ \citenamefont {Rigol}}]{Cazalilla:2011dma}%
  \BibitemOpen
  \bibfield  {author} {\bibinfo {author} {\bibfnamefont {M.~A.}\ \bibnamefont
  {Cazalilla}}, \bibinfo {author} {\bibfnamefont {R.}~\bibnamefont {Citro}},
  \bibinfo {author} {\bibfnamefont {T.}~\bibnamefont {Giamarchi}}, \bibinfo
  {author} {\bibfnamefont {E.}~\bibnamefont {Orignac}}, \ and\ \bibinfo
  {author} {\bibfnamefont {M.}~\bibnamefont {Rigol}},\ }\href {\doibase
  10.1103/RevModPhys.83.1405} {\bibfield  {journal} {\bibinfo  {journal} {Rev.
  Mod. Phys.}\ }\textbf {\bibinfo {volume} {83}},\ \bibinfo {pages} {1405}
  (\bibinfo {year} {2011})}\BibitemShut {NoStop}%
\bibitem [{\citenamefont {Carrasquilla}\ \emph {et~al.}(2013)\citenamefont
  {Carrasquilla}, \citenamefont {Manmana},\ and\ \citenamefont
  {Rigol}}]{Carrasquilla:2013dg}%
  \BibitemOpen
  \bibfield  {author} {\bibinfo {author} {\bibfnamefont {J.}~\bibnamefont
  {Carrasquilla}}, \bibinfo {author} {\bibfnamefont {S.~R.}\ \bibnamefont
  {Manmana}}, \ and\ \bibinfo {author} {\bibfnamefont {M.}~\bibnamefont
  {Rigol}},\ }\href {\doibase 10.1103/PhysRevA.87.043606} {\bibfield  {journal}
  {\bibinfo  {journal} {Phys. Rev. A}\ }\textbf {\bibinfo {volume} {87}},\
  \bibinfo {pages} {043606} (\bibinfo {year} {2013})}\BibitemShut {NoStop}%
\bibitem [{\citenamefont {Bo{\'e}ris}\ \emph {et~al.}(2015)\citenamefont
  {Bo{\'e}ris}, \citenamefont {Gori}, \citenamefont {Hoogerland}, \citenamefont
  {Kumar}, \citenamefont {Lucioni}, \citenamefont {Tanzi}, \citenamefont
  {Inguscio}, \citenamefont {Giamarchi}, \citenamefont {D'Errico},
  \citenamefont {Carleo}, \citenamefont {Modugno},\ and\ \citenamefont
  {Sanchez-Palencia}}]{Boeris:2015un}%
  \BibitemOpen
  \bibfield  {author} {\bibinfo {author} {\bibfnamefont {G.}~\bibnamefont
  {Bo{\'e}ris}}, \bibinfo {author} {\bibfnamefont {L.}~\bibnamefont {Gori}},
  \bibinfo {author} {\bibfnamefont {M.~D.}\ \bibnamefont {Hoogerland}},
  \bibinfo {author} {\bibfnamefont {A.}~\bibnamefont {Kumar}}, \bibinfo
  {author} {\bibfnamefont {E.}~\bibnamefont {Lucioni}}, \bibinfo {author}
  {\bibfnamefont {L.}~\bibnamefont {Tanzi}}, \bibinfo {author} {\bibfnamefont
  {M.}~\bibnamefont {Inguscio}}, \bibinfo {author} {\bibfnamefont
  {T.}~\bibnamefont {Giamarchi}}, \bibinfo {author} {\bibfnamefont
  {C.}~\bibnamefont {D'Errico}}, \bibinfo {author} {\bibfnamefont
  {G.}~\bibnamefont {Carleo}}, \bibinfo {author} {\bibfnamefont
  {G.}~\bibnamefont {Modugno}}, \ and\ \bibinfo {author} {\bibfnamefont
  {L.}~\bibnamefont {Sanchez-Palencia}},\ }\href
  {http://arxiv.org/abs/1509.04742} {\  (\bibinfo {year} {2015})},\ \Eprint
  {http://arxiv.org/abs/1509.04742} {1509.04742} \BibitemShut {NoStop}%
\bibitem [{\citenamefont {Astrakharchik}\ \emph {et~al.}(2015)\citenamefont
  {Astrakharchik}, \citenamefont {Krutitsky}, \citenamefont {Lewenstein},\ and\
  \citenamefont {Mazzanti}}]{Astrakharchik:2015ww}%
  \BibitemOpen
  \bibfield  {author} {\bibinfo {author} {\bibfnamefont {G.~E.}\ \bibnamefont
  {Astrakharchik}}, \bibinfo {author} {\bibfnamefont {K.~V.}\ \bibnamefont
  {Krutitsky}}, \bibinfo {author} {\bibfnamefont {M.}~\bibnamefont
  {Lewenstein}}, \ and\ \bibinfo {author} {\bibfnamefont {F.}~\bibnamefont
  {Mazzanti}},\ }\href {http://arxiv.org/abs/1509.01424} {\  (\bibinfo {year}
  {2015})},\ \Eprint {http://arxiv.org/abs/1509.01424} {1509.01424}
  \BibitemShut {NoStop}%
\bibitem [{\citenamefont {Aharonov}\ and\ \citenamefont
  {Susskind}(1967)}]{Aharonov:1967be}%
  \BibitemOpen
  \bibfield  {author} {\bibinfo {author} {\bibfnamefont {Y.}~\bibnamefont
  {Aharonov}}\ and\ \bibinfo {author} {\bibfnamefont {L.}~\bibnamefont
  {Susskind}},\ }\href {\doibase 10.1103/PhysRev.155.1428} {\bibfield
  {journal} {\bibinfo  {journal} {Phys. Rev.}\ }\textbf {\bibinfo {volume}
  {155}},\ \bibinfo {pages} {1428} (\bibinfo {year} {1967})}\BibitemShut
  {NoStop}%
\bibitem [{\citenamefont {Herdman}\ \emph {et~al.}(2014)\citenamefont
  {Herdman}, \citenamefont {Inglis}, \citenamefont {Roy}, \citenamefont
  {Melko},\ and\ \citenamefont {Del~Maestro}}]{Herdman:2014ey}%
  \BibitemOpen
  \bibfield  {author} {\bibinfo {author} {\bibfnamefont {C.~M.}\ \bibnamefont
  {Herdman}}, \bibinfo {author} {\bibfnamefont {S.}~\bibnamefont {Inglis}},
  \bibinfo {author} {\bibfnamefont {P.~N.}\ \bibnamefont {Roy}}, \bibinfo
  {author} {\bibfnamefont {R.~G.}\ \bibnamefont {Melko}}, \ and\ \bibinfo
  {author} {\bibfnamefont {A.}~\bibnamefont {Del~Maestro}},\ }\href {\doibase
  10.1103/PhysRevE.90.013308} {\bibfield  {journal} {\bibinfo  {journal} {Phys.
  Rev. E}\ }\textbf {\bibinfo {volume} {90}},\ \bibinfo {pages} {013308}
  (\bibinfo {year} {2014})}\BibitemShut {NoStop}%
\bibitem [{\citenamefont {Wootters}(1998)}]{Wootters1998}%
  \BibitemOpen
  \bibfield  {author} {\bibinfo {author} {\bibfnamefont {W.~K.}\ \bibnamefont
  {Wootters}},\ }\href {\doibase 10.1103/PhysRevLett.80.2245} {\bibfield
  {journal} {\bibinfo  {journal} {Phys. Rev. Lett.}\ }\textbf {\bibinfo
  {volume} {80}},\ \bibinfo {pages} {2245} (\bibinfo {year}
  {1998})}\BibitemShut {NoStop}%
\bibitem [{\citenamefont {Vidal}\ and\ \citenamefont
  {Werner}(2002)}]{Vidal2002}%
  \BibitemOpen
  \bibfield  {author} {\bibinfo {author} {\bibfnamefont {G.}~\bibnamefont
  {Vidal}}\ and\ \bibinfo {author} {\bibfnamefont {R.~F.}\ \bibnamefont
  {Werner}},\ }\href {\doibase 10.1103/PhysRevA.65.032314} {\bibfield
  {journal} {\bibinfo  {journal} {Phys. Rev. A}\ }\textbf {\bibinfo {volume}
  {65}},\ \bibinfo {pages} {032314} (\bibinfo {year} {2002})},\ \Eprint
  {http://arxiv.org/abs/0102117} {0102117} \BibitemShut {NoStop}%
\bibitem [{\citenamefont {Osterloh}\ \emph {et~al.}(2002)\citenamefont
  {Osterloh}, \citenamefont {Amico}, \citenamefont {Falci},\ and\ \citenamefont
  {Fazio}}]{Osterloh2002}%
  \BibitemOpen
  \bibfield  {author} {\bibinfo {author} {\bibfnamefont {A.}~\bibnamefont
  {Osterloh}}, \bibinfo {author} {\bibfnamefont {L.}~\bibnamefont {Amico}},
  \bibinfo {author} {\bibfnamefont {G.}~\bibnamefont {Falci}}, \ and\ \bibinfo
  {author} {\bibfnamefont {R.}~\bibnamefont {Fazio}},\ }\href {\doibase
  10.1038/416608a} {\bibfield  {journal} {\bibinfo  {journal} {Nature}\
  }\textbf {\bibinfo {volume} {416}},\ \bibinfo {pages} {608} (\bibinfo {year}
  {2002})}\BibitemShut {NoStop}%
\bibitem [{\citenamefont {Osborne}\ and\ \citenamefont
  {Nielsen}(2002)}]{Osborne2002}%
  \BibitemOpen
  \bibfield  {author} {\bibinfo {author} {\bibfnamefont {T.}~\bibnamefont
  {Osborne}}\ and\ \bibinfo {author} {\bibfnamefont {M.}~\bibnamefont
  {Nielsen}},\ }\href {\doibase 10.1103/PhysRevA.66.032110} {\bibfield
  {journal} {\bibinfo  {journal} {Phys. Rev. A}\ }\textbf {\bibinfo {volume}
  {66}},\ \bibinfo {pages} {032110} (\bibinfo {year} {2002})}\BibitemShut
  {NoStop}%
\bibitem [{\citenamefont {Fr{\'e}rot}\ and\ \citenamefont
  {Roscilde}(2015)}]{Frerot:2015tl}%
  \BibitemOpen
  \bibfield  {author} {\bibinfo {author} {\bibfnamefont {I.}~\bibnamefont
  {Fr{\'e}rot}}\ and\ \bibinfo {author} {\bibfnamefont {T.}~\bibnamefont
  {Roscilde}},\ }\href {http://arxiv.org/abs/1512.00805} {\  (\bibinfo {year}
  {2015})},\ \Eprint {http://arxiv.org/abs/1512.00805} {1512.00805}
  \BibitemShut {NoStop}%
\bibitem [{\citenamefont {Horodecki}\ \emph {et~al.}(1996)\citenamefont
  {Horodecki}, \citenamefont {Horodecki},\ and\ \citenamefont
  {Horodecki}}]{Horodecki1996a}%
  \BibitemOpen
  \bibfield  {author} {\bibinfo {author} {\bibfnamefont {M.}~\bibnamefont
  {Horodecki}}, \bibinfo {author} {\bibfnamefont {P.}~\bibnamefont
  {Horodecki}}, \ and\ \bibinfo {author} {\bibfnamefont {R.}~\bibnamefont
  {Horodecki}},\ }\href {\doibase 10.1016/S0375-9601(96)00706-2} {\bibfield
  {journal} {\bibinfo  {journal} {Phys. Lett. A}\ }\textbf {\bibinfo {volume}
  {223}},\ \bibinfo {pages} {1} (\bibinfo {year} {1996})}\BibitemShut {NoStop}%
\bibitem [{\citenamefont {Horodecki}\ \emph {et~al.}(1997)\citenamefont
  {Horodecki}, \citenamefont {Horodecki},\ and\ \citenamefont
  {Horodecki}}]{Horodecki1997}%
  \BibitemOpen
  \bibfield  {author} {\bibinfo {author} {\bibfnamefont {M.}~\bibnamefont
  {Horodecki}}, \bibinfo {author} {\bibfnamefont {P.}~\bibnamefont
  {Horodecki}}, \ and\ \bibinfo {author} {\bibfnamefont {R.}~\bibnamefont
  {Horodecki}},\ }\href {\doibase 10.1103/PhysRevLett.78.574} {\bibfield
  {journal} {\bibinfo  {journal} {Phys. Rev. Lett.}\ }\textbf {\bibinfo
  {volume} {78}},\ \bibinfo {pages} {574} (\bibinfo {year} {1997})}\BibitemShut
  {NoStop}%
\end{thebibliography}%

\end{document}